\documentstyle[twocolumn,prl,aps,psfig,amssymb]{revtex}
\newcommand{\equ}{Eq.}

\newcommand{\fig}{Fig.}
\newcommand{\figs}{Figs.}

\newcommand{\rem}[1]{}

\newcommand{\bm}[1]{\mbox{\boldmath$#1$\unboldmath}}

\newtheorem{platz}{{Fig.}} 
\newcommand{\FIGo}[3]{\begin{figure}%
#3%
\caption[]{\footnotesize #2}%
\label{#1}%
\end{figure}}

\rem{
\newcommand{\FIGo}[3]{%
\marginpar{\begin{platz} \label{#1} ~ \end{platz} \vspace*{1.5ex} }
}
}

\newcommand{\r}{\rho}
\newcommand{\mean}[1]{\langle #1\rangle}
\newcommand{\const}{\mbox{constant}}
\newcommand{\mod}{\,{\rm mod}\,1}

\begin{document}
 \twocolumn[\hsize\textwidth\columnwidth\hsize\csname @twocolumnfalse\endcsname
\draft
\begin{title}
 { Devil's Staircase in Magnetoresistance of a Periodic Array of Scatterers}
\end{title}
\author{Jan Wiersig and Kang-Hun Ahn}
\address{Max-Planck-Institut f\"ur Physik komplexer Systeme, 
N\"othnitzer Str. 38, 01187 Dresden, Germany      \\ }
\date{\today}
\maketitle

\widetext

\begin{abstract}
The nonlinear response to an external electric field is studied for classical
non-interacting charged particles under the influence of a uniform
magnetic field, a periodic potential, and an effective friction force.
We find numerical and analytical evidence that the ratio of transversal to
longitudinal resistance forms a Devil's staircase. 
The staircase is attributed to the dynamical phenomenon of mode-locking.
\vspace{-0.5cm}
\end{abstract}
]
\narrowtext

The electron transport in two-dimensional periodic arrays of scatterers
has been actively studied for the last decade.
One of the most interesting feature is the plateau-like behaviour in
Hall resistance as well as peaks in magnetoresistance\cite{ensslin,weiss}
at low magnetic fields below the quantum Hall regime.
The peak structure in magnetoresistance has been 
attributed to the electron cyclotron orbits which enclose an {\it integer}
number of scatterers~\cite{weiss,FGK92}.
The cyclotron motion is important when
the electron mean free path $l$ 
($l$ is measured in the absence of periodic scatterers)
is greater than the period $a$ of the regular scatterers. 

In this Letter, we present a theory for electron transport in the
deeply {\it diffusive} regime $l \ll a$ where cyclotron motion is not relevant. 
We predict a new and interesting effect where the structure in the
magnetoresistance is associated with {\it fractional} numbers.  
We will show that the relative ratio between the transversal
and longitudinal magnetoresistance forms a {\it Devil's
staircase}. These fractal staircases originate from a 
dynamical phenomenon known as {\it mode-locking} which appears naturally in the
context of {\it circle maps}~\cite{AP90}. 
Our prediction is based on the finding that the dynamics in a two-dimensional
periodic array of scatterers with intrinsic momentum relaxation effectively
reduces to a circle map.  

Let us consider first a particle of mass $m$, charge $q$ in crossed electric
(${\bm E} = E\hat{\bm x}$) and magnetic fields (${\bm B} = -B\hat{\bm z}$) with
intrinsic momentum relaxation.
The resulting drift motion ${\bm x(t)} = (x(t),y(t))$ can be described by
adding a frictional force, proportional to the drift velocity ${\bm v}$, to the
equations of motion yielding 
\begin{eqnarray}\label{eq:eqm}
\nonumber
\dot{x} = v_x & , & \quad m\dot{v}_x = -q B v_y + q E - m v_x/\tau \\
\dot{y} = v_y & , & \quad m\dot{v}_y = q B v_x - m v_y/\tau 
\label{eq_motion}
\end{eqnarray}
with $\tau$ being the momentum relaxation time. Cyclotron motion only exists
for a transient time; each trajectory finally converges to a straight line
${\bm x}(t\to\infty) = {\bm x}(0)+\mean{{\bm v}}t$, with the time-averaged 
velocity $\mean{{\bm v}} = (m/\tau,qB)qE/(q^2B^2+m^2/\tau^2)$. 
With the charge density $n$ and the current density ${\bm J} = qn\mean{{\bm
v}}$, it is easy to calculate the resistivities defined by ${\bm E} =
\hat{\rho}{\bm J}$ and Onsager's relations $\rho_{xx} = \rho_{yy}$ and
$\rho_{yx} = -\rho_{xy}$: the diagonal resistivity $\rho_{xx} = m/q^2n\tau$ is
constant, whereas the off-diagonal resistivity $\rho_{xy} = B/qn$ depends
linearly on the magnetic field -- no plateaus are present.  
 
We now add a periodic potential $V=V_0F({\bm x})$ with $|F|$ of order
one to the equations of motion~(\ref{eq:eqm}), assuming that the
periodicity of the potential, $a$, is much larger than the length
scale related to the intrinsic momentum relaxation.    
When we scale the coordinates ${\bm x}$ with $a$ and the time $t$ with 
$\tau$, we get the same equations of motion for a given set of three
dimensionless parameters: $\tilde{B} = q\tau B/m$,
$\tilde{E}=q\tau^{2}E/(ma)$, and $\tilde{V}_0 = V_{0}/(qEa)$. 
We use the rescaled magnetic field $\tilde{B}$ as a variable parameter
and fix the rescaled electric field and potential strength to
$\tilde{E}=15.3$ and $\tilde{V}_0=0.12$ if not otherwise stated.
Though we could choose various sets of parameters ($\tau, a, E$, etc.) for 
the chosen values of $\tilde{E}$ and $\tilde{V}_0$, for definiteness
we will take $\tau=7.6~10^{-14} s$, 
$E=-2.3 ~10^{4} V/cm$, and $a=0.5\mu m$, and fix these for the rest of the
paper.  
The particle is an electron in GaAs sample 
with $q=-e$ and $m=0.067 m_{e}$. 
By taking a typical value of Fermi velocity $v_{f}\approx 3 \times 10^{7} cm/s$,
the mean free path of our system is 
$l=v_{f}\tau\approx 0.02 \mu m \ll a$.

First, we show how the particle's motion is related to a one-dimensional circle
map through analytical considerations for sufficiently small potential
strength. Note that our numerics do not rely on this requirement.  
Due to the periodicity of the potential it is sufficient to consider the unit
cell of the potential with periodic boundary conditions, i.e. we can
compactify the $(x,y)$-plane to a two-dimensional torus. A typical asymptotic
solution for zero potential strength as discussed above appears here as a
quasiperiodic trajectory filling densely a two-dimensional invariant torus
${\bm v } = \const$ in four-dimensional phase space. Trajectories starting
away from the torus are attracted towards it and asymptotically converge onto
it.  
For finite potential strength the torus is smoothly deformed or collapsed into
lower-dimensional objects or broadened to a higher-dimensional object (loosely
speaking, a torus with finite but very small thickness). 
The latter possibility can be ignored to a very good approximation as we will
see later. 
A point on the torus (and on the lower-dimensional objects) is uniquely
labelled by unit-cell coordinates $(x_1\mod,x_2\mod)$ defined by ${\bm x}
= x_1 {\bm a}_1 + x_2 {\bm a}_2$, with ${\bm a}_1 = (a_{11},a_{12})$, ${\bm
a}_2 = (a_{21},a_{22})$ being the
lattice vectors of the unit cell. The long-time behavior is therefore
completely described by two-dimensional dynamics $[x_1(t),x_2(t)]$. 
As usual, such dynamics can be conveniently investigated by introducing {\it
Poincar\'e surfaces of section} defined by $x_1(t_n)\mod = 0$, where $n$ is an
integer and modulo 1 restricts the variable to the interval $[0,1)$. 

The discrete evolution of $x_{2,n} \equiv  x_2(t_n)$ is then governed by a
one-dimensional map $x_{2,n+1} = f(x_{2,n})$ where 
$f$ is a smooth and monotonic function. Due to the periodicity of the
potential the condition $f(x_2)\mod = f(x_2\mod)$ holds. 
Such a map is called an {\it invertible circle map}. Its {\it rotation
number},   
\begin{equation}\label{eq:rho}
\r \equiv \lim_{n \to \infty} \frac{x_{2,n} - x_{2,0}}{n} \ ,
\end{equation}
is well-defined and independent of the initial condition
$x_{2,0}$~\cite{AP90}. A rational value of $\r=p/q$ ($p$ and $q$ are
integers without common divisor) indicates periodic motion,
whereas an irrational value indicates quasiperiodic motion; chaotic motion is
not possible.

We now derive a simplified circle map valid in the limit $\tilde{V}_0 \ll 1$.  
If $\tilde{V}_0=0$ then the asymptotic dynamics obeys $\dot{v}_x=\dot{v}_y=0$,
leading to $f(x_2) = x_2+\r_0$ with the constant $\r_0(\tilde{B})$; here we
have simply $\r=\r_0$, so $\r_0$ can be regarded as the unperturbed rotation
number. 
The asymptotic dynamics for nonzero but sufficiently small $\tilde{V}_0$ is
overdamped, i.e. $|\dot{v}_x| \ll |v_x|/\tau$ and $|\dot{v}_y| \ll
|v_y|/\tau$. As in the potential-free case the number of differential
equations reduces to two, enabling us to determine $f$ up to first order in
$\tilde{V}_0$  
\begin{equation}\label{eq:f}
 f(x_{2,n};\r_0,\tilde{E},\tilde{V}_0) \approx 
x_{2,n}+\r_0+\tilde{V}_0\; g(x_{2,n};\r_0) 
\label{map}
\end{equation}
with 
\begin{eqnarray}
g & = & \frac{\r_0^2(a_{22}^2+a_{21}^2)+2\r_0(a_{11}a_{21}+a_{12}a_{22})+a_{12}^2+a_{11}^2}{(a_{11}a_{22}-a_{21}a_{12})^2} \times
\nonumber\\
& & \int_0^1\left(a_{21}\frac{\partial F}{\partial
x_1}-a_{11}\frac{\partial F}{\partial x_2}\right)\Big|_{x_2=\r_0x_1+x_{2,n}}
dx_1 
\ .
\end{eqnarray}
$g$ is a nontrivial periodic function of $x_{2,n}$. Note that $f$ is
independent of the parameter $\tilde{E}$ in the present approximation.
Using Stoke's theorem it can be shown that $\int_0^1gdx_2 = 0$. 
For circle maps of the form given in \equ~(\ref{eq:f}) with the above-mentioned
properties of $g$ it is proven that for each
$\tilde{V}_0$ there exists a Devil's staircase, a monotonically increasing
function $\r = \r(\r_0)$ with plateaus around each rational value of
$\r_0$. The width of the plateaus, $\Delta \r_0$, is proportional to
$\tilde{V}_0$~\cite{Hall84}. 
Similarly, the widths of the plateaus, $\Delta \tilde{B}$, of the function
$\r=\r(\tilde{B})$, are proportional to $\tilde{V}_0$ since
$\Delta \tilde{B} \approx (d\tilde{B}/d\r_0)\Delta\r_0$.  
  
The rotation number $\r$ is directly related to the continuous-time
dynamics. This can be seen by rewriting \equ~(\ref{eq:rho}) with ${\bm v}  =
v_1 {\bm a}_1 + v_2 {\bm a}_2$ as    
\begin{equation}\label{eq:rn0}
\r = \frac{\mean{v_2}}{\mean{v_1}} = 
\frac{-a_{12}+a_{11}\mean{v_y}/\mean{v_x}}{a_{22}-a_{21}\mean{v_y}/\mean{v_x}}
= 
\frac{-a_{12}+a_{11}\sigma_{xy}/\sigma_{xx}}{a_{22}-a_{21}\sigma_{xy}/\sigma_{xx}}
\ .
\end{equation}
Hence, $\r$ is given by the ratio of the conductivities. In the case that the periodic potential is
invariant under $\pi/m$-rotations, $m=2,3,\ldots$, Onsager's
relations show that $\r$ can also be expressed by the ratio
$\rho_{xy}/\rho_{xx}$. 

%
We now present numerical evidence showing that for finite $\tilde{V}_0 \leq 0.4$ the
continuous-time dynamics is indeed given by a one-dimensional circle map and
that the magnetoresistance shows a fractal structure. 
As an example of the periodic scatterers, we take a potential 
with hexagonal symmetry shown in \fig~\ref{fig:potorb}a,  
\begin{eqnarray}\label{eq:potential}
\nonumber
V(x,y) & = &
V_0\left\{\left[\cos\left(\frac{\pi}{3}
(\sqrt{3}x-y)\right)\cos\left(\frac{2\pi}{3}y\right)\right]^\beta
\right.
\\
\nonumber
& + & \left[\cos\left(\frac{\pi}{3}
(\sqrt{3}x+y)\right)\cos\left(\frac{\pi}{3}(\sqrt{3}x-y)\right)\right]^\beta
\\
& + & \left.
\left[\cos\left(\frac{2\pi}{3}y\right)\cos\left(\frac{\pi}{3}
(\sqrt{3}x+y)\right)\right]^\beta
\right\} \ ,
\end{eqnarray}
with $\beta=8$. The lattice vectors are ${\bm a}_1 = (\sqrt{3},0)$ and ${\bm
a}_2 = (-\sqrt{3}/2,-3/2)$.   
We have used the Runge-Kutta method~\cite{Press88} with $10^6$ time steps
of variable size to integrate the equations of motion. We found that all
initial conditions, for fixed parameters, lead to the same mean velocity
$\mean{{\bm v}}$. So only one single orbit is needed to calculate the current
density and the resistivities without employing the Kubo
formula~\cite{Kubo57}. This already indicates the collapse to an invertible 
circle map. 
To show the reduction to the map explicitly, we calculate a sequence of points 
$(x_{2,n}\mod,x_{2,n+1}\mod)$ by solving the complete set of 
differential equations for different initial velocities from one Poincar\'e
section to the next. 
Figures~\ref{fig:fg}a and b show that these points indeed lie very close to a
line, so $x_{2,n+1}$ is to excellent approximation
a function of $x_{2,n}$ independent of the initial velocity.
We have computed $(x_{2,n+1} - x_{2,n}-\r_0)/\tilde{V}_0$ for different
parameters $\tilde{V}_0$ and $\tilde{E}$. It can be seen from
\fig~\ref{fig:fg}c that this quantity is roughly independent of $\tilde{V}_0$,
even though small deviations indicate weak nonlinear behavior in
$\tilde{V}_0$. Figure~\ref{fig:fg}d shows that $(x_{2,n+1} -
x_{2,n}-\r_0)/\tilde{V}_0$ does not change when $\tilde{E}$ is varied over 
two orders of magnitude.  
Figures~\ref{fig:fg}a-d therefore confirm the reduction to the circle map and
its parameter dependence as predicted in \equ~(\ref{eq:f}).
\def\figpotorb{%
a) Contourplot of the potential in \equ~(\ref{eq:potential}) with $\beta=8$.
b) Orbit with $B = 8.7T$ and $\r=1$ in the unit cell.  
c) $B = 9.77T$ and $\r=1$.
d) $9.8T$ and $\r=1.01407...$.
e) $2.9T$ and $\r=1/2$.
f) $4.4T$ and $\r=2/3$.
}
\def\FIGpotorb{\centerline{\psfig{figure=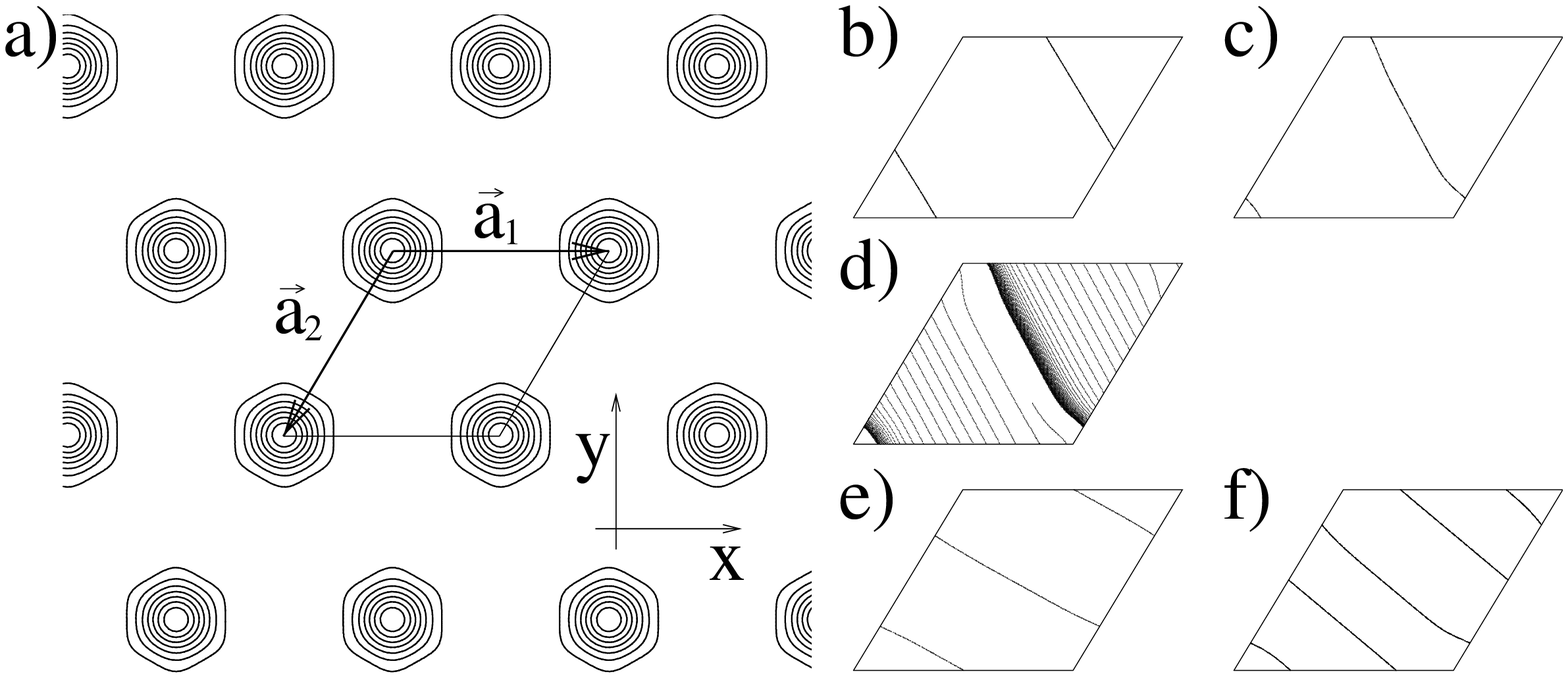,width=8.0cm,angle=0}
\vspace{0.2cm}
}}
\FIGo{fig:potorb}{\figpotorb}{\FIGpotorb}
\def\figfg{%
a) $x_{2,n+1}\mod$ vs. $x_{2,n}$ for $B = 8.7T$.
b) $9.77T$.
c) $(x_{2,n+1} - x_{2,n}-\r_0)/\tilde{V}_0$ vs. $x_{2,n}$ for $B = 8.7T$,
$\tilde{V}_0=0.12$ and $\tilde{E}=15.3$ (black); $\tilde{V}_0=0.06$
(red); $\tilde{V}_0=0.03$ (green).
d) $\tilde{V}_0=0.12$ and $\tilde{E}=15.3$ (black);
$\tilde{E}=1.53$ (red); $\tilde{E}=0.153$ (green).
} 
\def\FIGfg{\centerline{\psfig{figure=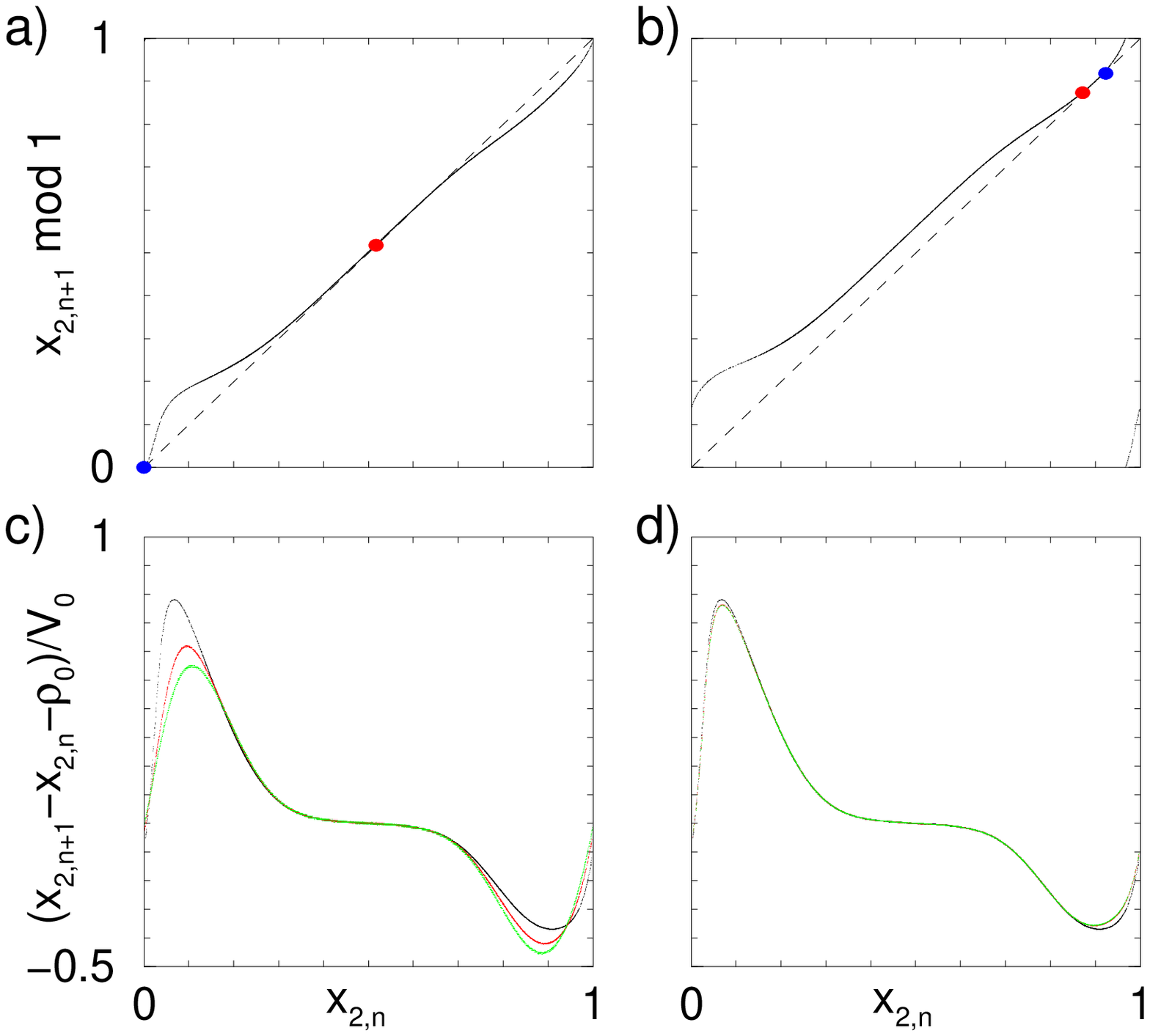,width=8.0cm,angle=0}
\vspace{0.1cm}
}}
\FIGo{fig:fg}{\figfg}{\FIGfg}

Figure~\ref{fig:staircase} is our main result. It shows the Devil's staircase
$\r$ vs. $\r_0$. Both quantities are computed from the resistivities via the
relations $\r = 2/(1+\sqrt{3}\rho_{xx}/\rho_{xy})$ and $\r_0 =
2/(1-\sqrt{3}/\tilde{B})$ from \equ~(\ref{eq:rn0}) with the lattice vectors of
the rotational symmetric potential~(\ref{eq:potential}).   
Note the exactness of the plateaus. 
Let us discuss some features of the Devil's staircase in relation
to the continuous and discrete-time dynamics.
Figures~\ref{fig:potorb}b and c show two orbits moving
in slightly different magnetic fields. Both have synchronized 
velocities $\mean{v_1}=\mean{v_2}$, i.e. $\r = 1$.  
Each orbit with $\r=1$ is a stable fixed point $x_2^*=f(x_2^*)$ in the
corresponding map (red circles in \figs~\ref{fig:fg}a and b; blue circles mark
unstable fixed points). This synchronization phenomenon, usually called {\it
mode-locking} in nonlinear dynamics, is therefore due to the robustness of
fixed points under variation of a parameter, here $\r_0(B)$.
Orbits with $\r \neq 1$ are in a different {\it mode}; in particular, orbits
with irrational rotation number are not periodic. In \fig~\ref{fig:potorb}d we
see an orbit with 
irrational $\r$ close to one. It stays for a long time near a periodic orbit
with $\r=1$ but from time to time it escapes, thereby filling the entire unit
cell densely. This phenomenon is called {\it intermittency}~\cite{Ott93}.  
In terms of the circle map, the transition from $\r=1$
to irrational $\r$ is a saddlenode bifurcation: as $B$ is varied,
stable and unstable fixed points come closer and closer as illustrated in 
\figs~\ref{fig:fg}a and b, and finally destroy each other, yielding
quasiperiodic motion.  
Varying $B$ further can lead again to periodic
motion. Figures~\ref{fig:potorb}e and f give two examples with $\r = 1/2$ and
$\r = 2/3$. The corresponding map has a stable fixed point of higher period,
e.g. for $\r = 1/2$ the fixed point condition is $x_2^*=f[f(x_2^*)]$. 
\def\figstaircase{%
Rotation number~$\r$ vs. $\r_0$. Arrows assign the rational value of the
rotation number $\r=p/q$ to each plateau.
The inset highlights the fractal structure of
the Devil's staircase.}
\def\FIGstaircase{\centerline{\psfig{figure=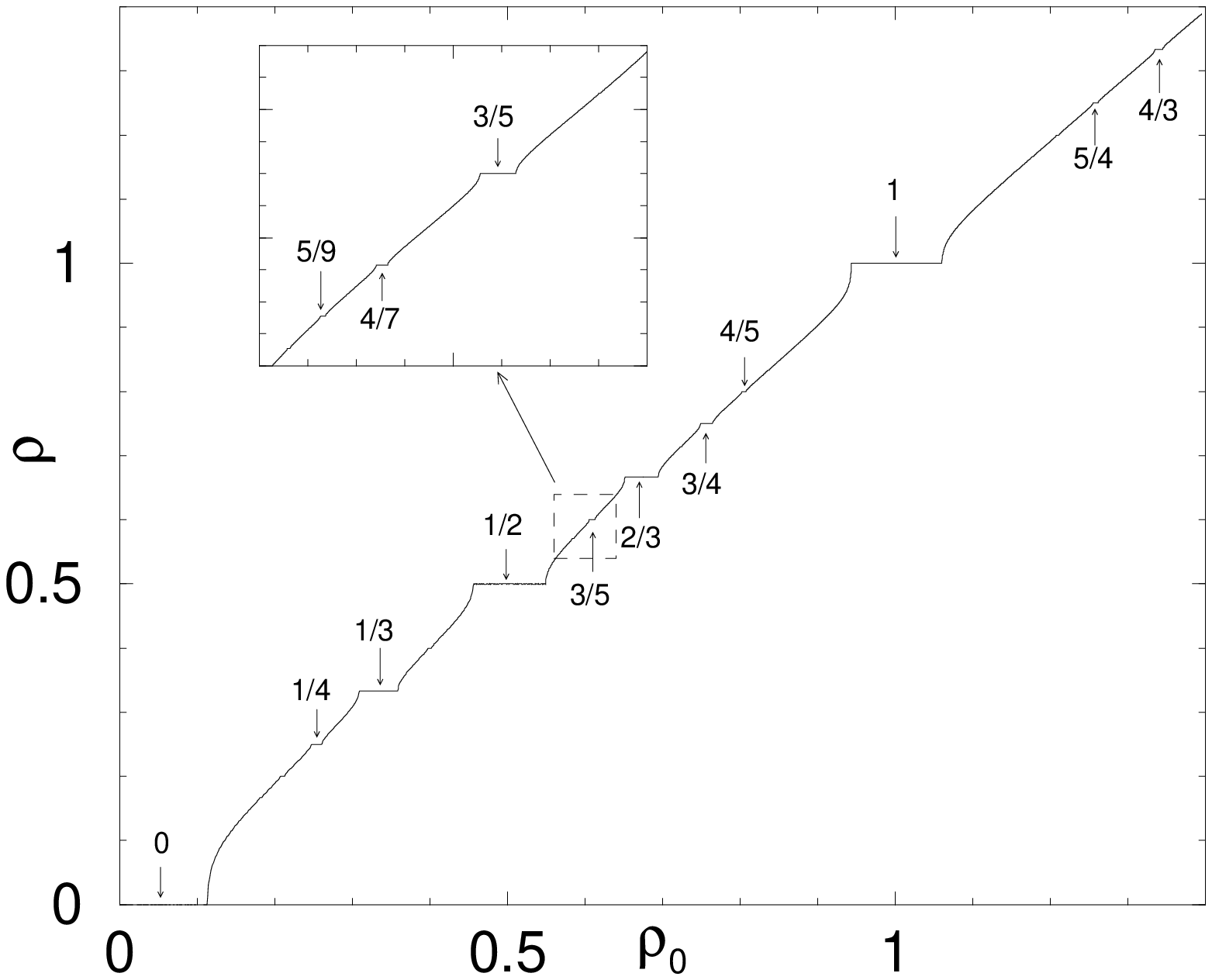,width=8.0cm,angle=0}
\vspace{0.2cm}
}}
\FIGo{fig:staircase}{\figstaircase}{\FIGstaircase}

We also observe a Devil's staircase-like function $\rho_{xy}(B)$ in
\fig~\ref{fig:res}, which reminds one of the quantum Hall
effects~\cite{klitzing,tsui}, but here the `plateaus' are not perfectly flat. 
One may note also that the way $\rho_{xx}$ varies for a number of
$\rho_{xy}$-`plateaus' has similarities to the quantum Hall
effects: going from the centre of a $\rho_{xy}$-`plateau'
towards its border, $\rho_{xx}$ increases and reaches its maximum exactly when
the `plateau' ends.  
This can be explained qualitatively. First, note that an orbit from 
the centre of the $\r=1$-plateau (\fig~\ref{fig:potorb}b) avoids the
potential's steep maxima at the corners of the unit cell. Hence, the
diagonal resistivity roughly equals the potential-free value $\rho_{xx} =
m/q^2n\tau \approx 3.12 k\Omega$.   
We infer from \figs~\ref{fig:potorb}b and c that the closer $B$ is to the
border of the plateau the closer is the orbit to the
potential's maxima. This effectively slows down the orbit, which explains the
increase of $\rho_{xx}$ (and $\rho_{xy}$).  
Leaving the plateau finally reduces the resistivities because the
intermittency effect becomes less pronounced with increasing $|\r-1|$; see
\fig~\ref{fig:potorb}d.   
\def\figres{%
Off-diagonal resistivity $\rho_{xy}$ and diagonal resistivity $\rho_{xx}$
vs. magnetic field $B$.  Arrows assign the rational value of the
rotation number $\r=p/q$ to each plateau.
}
\def\FIGres{\centerline{\psfig{figure=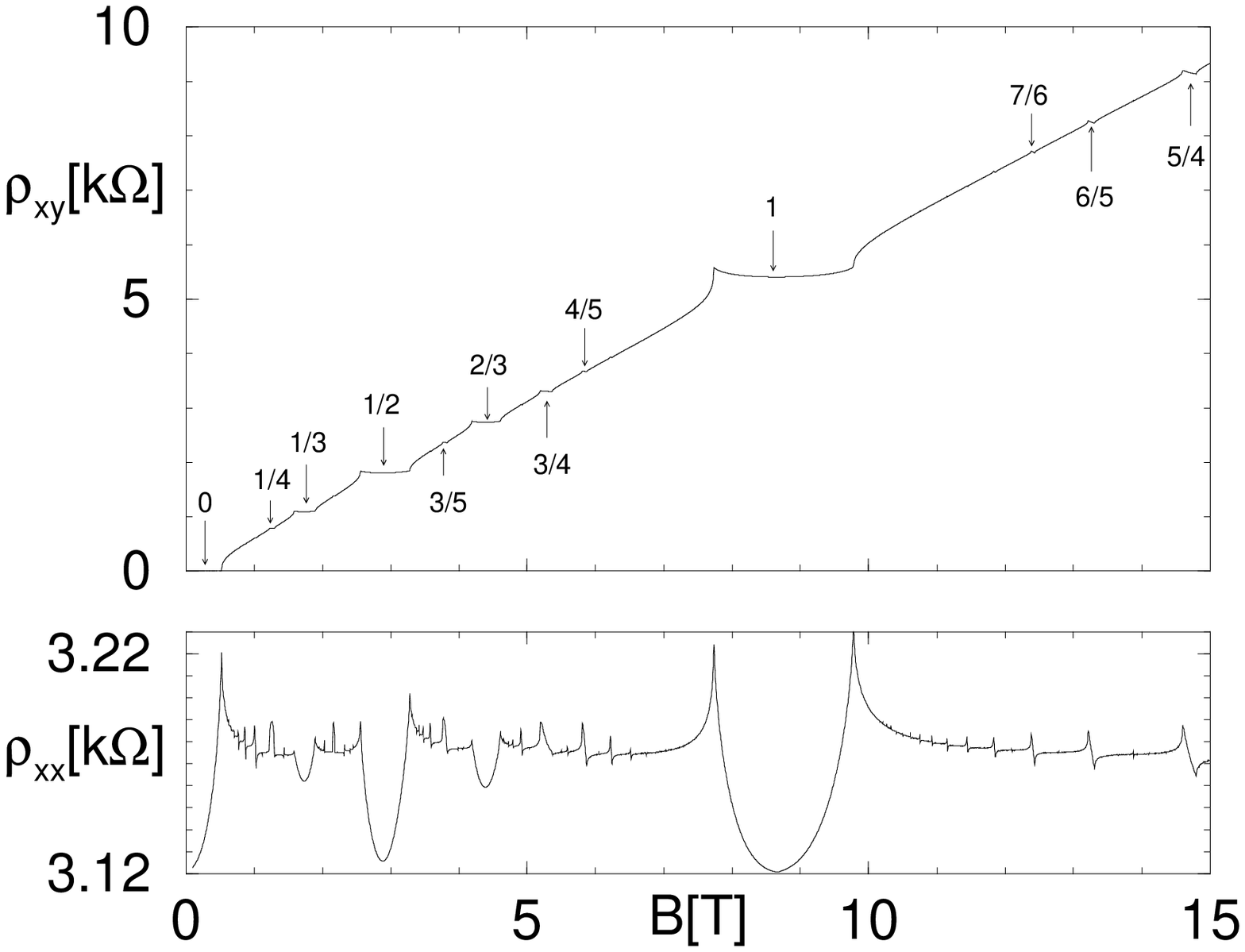,width=8.0cm,angle=0}
\vspace{0.2cm}
}}
\FIGo{fig:res}{\figres}{\FIGres}

The example in \fig~\ref{fig:plateau} confirms that the widths of the
plateaus, $\Delta B$, is roughly proportional to $\tilde{V}_0$ as predicted
from the analysis of the simplified circle map in \equ~(\ref{map}).
By noting that $\tilde{V_{0}}$ denotes the dimensionless potential strength we
see an interesting analogy with the quantum Hall effect where the size of
plateaus is sensitive to the disorder strength of the systems~\cite{janssen}.
Here we have the advantage of the use of the 
simplified circle map to analyse the plateau size, although 
the overall structure of the hierarchy of the plateaus is
also too complicated to be dealt with in an analytic fashion.
\def\figplateau{%
Dependence of the $\r=1$-plateau on $\tilde{V}_0$;
cf. \fig~\ref{fig:res}.} 
\def\FIGplateau{\centerline{\psfig{figure=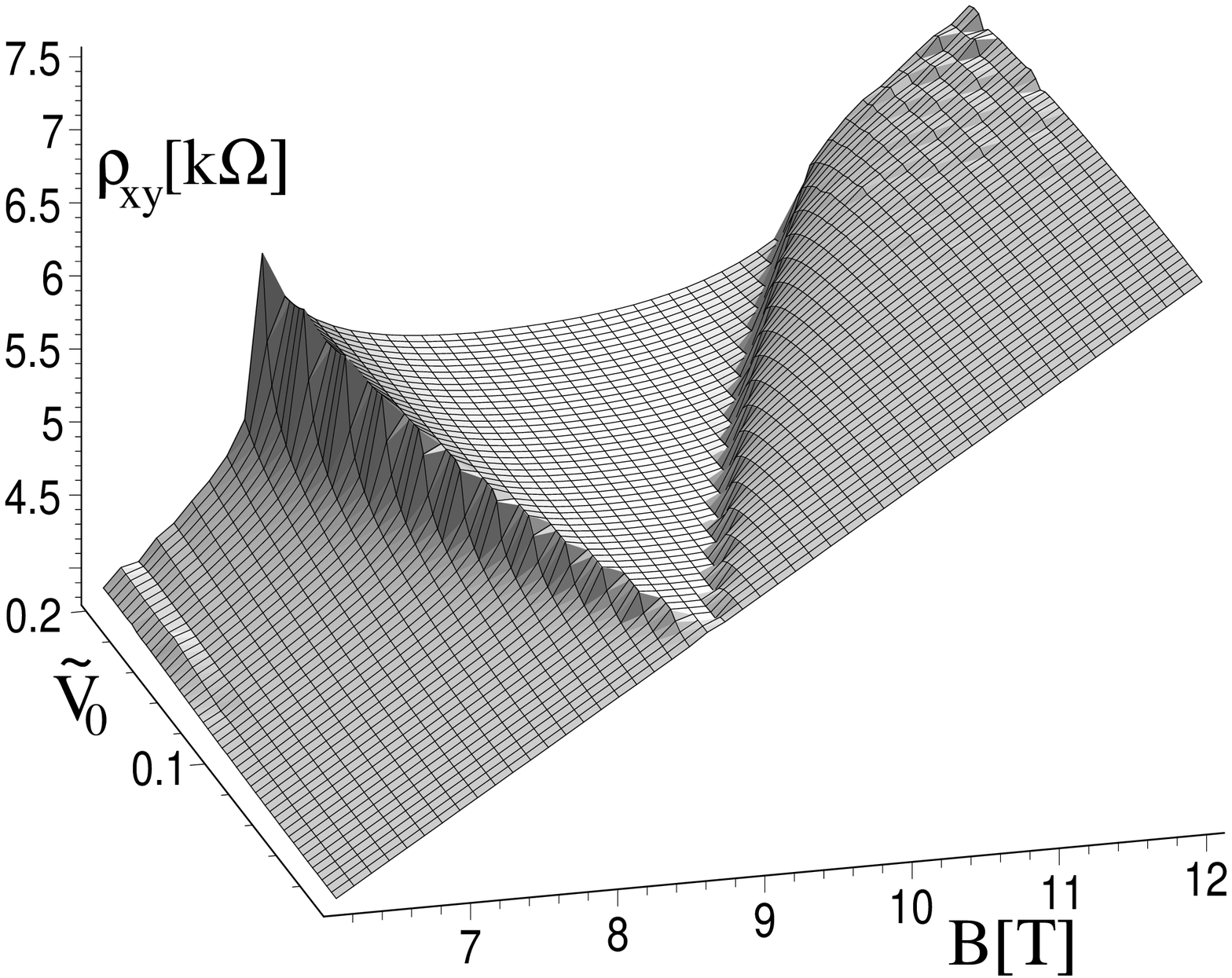,width=8.0cm,angle=0}
\vspace{0.2cm}
}}
\FIGo{fig:plateau}{\figplateau}{\FIGplateau}

Since the particle can be trapped in the very flat local minima of the
potential~(\ref{eq:potential}),  
a finite dc electric field is necessary to overcome the potential
trap as in the experiment in Ref.\cite{gusev}.
The threshold\cite{comment1} value $E_T$ of the electric field
for the finite electrical current might be crudely estimated
by $E_{T} \sim V_{0}/(qa)$, which is in agreement with our calculations
(not shown).  
However, in reality the threshold electric field will be much smaller
in the presence of the degenerate electron gas because the weak local
trap potential will be screened\cite{privat1}.
%
While we have chosen in our numerical calculation
a large electric field ($2.3~10^{4}$V/cm)
and a low mobility of the sample
 ($\mu=e\tau /m \approx 20000cm^{2}/Vs$),
the Devil's staircase may be observed 
in a broad range of parameters.
We also note that the specific form of the potential~(\ref{eq:potential}) is
not relevant. However, the steepness of the potential influences the shape of
the plateaus. Large steepness (large $\beta$) results in wide and curved
plateaus.   
It should be mentioned also that the Devil's staircase will not be seen
in the conventional Hall bar where the electrical current is fixed 
along the bar and the induced Hall voltage is measured.
To measure the electrical current in two-dimensional samples for fixed applied
voltages, one needs to use metallic leads which span the entire length of two
opposite edges of the sample~\cite{privat1}. 
 
%


In summary, we have calculated the magnetoresistance of a lateral surface
superlattice with strong momentum relaxation where the cyclotron motion is not
involved. Our calculations show a fractal plateau structure in the
magnetoresistance which stems from purely classical nonlinear dynamics. 
We have explained our calculational results in terms of the theory of circle
maps where the Devil's staircase is already well understood.

We acknowledge helpful discussions with P.~Fulde, K.~v.~Klitzing, R.~Gerhardts,
K.~Richter and R.~Klages.


\rem{
\bibliographystyle{prsty}
\bibliography{../bib/fg4,../bib/extern}
}

\end{document}